\documentclass[aps,prb,reprint,superscriptaddress,amsmath,amssymb]{revtex4-2}
\usepackage{graphicx}
\usepackage{bm}
\usepackage{color}

\newcommand{\kB}{k_{\text{\tiny B}}}

\begin{document}


\title{Acoustic phonon softening and lattice instability driven by on-site $f$-$d$ hybridization in CeCoSi}


\author{Takeshi Matsumura}
\email[]{tmatsu@hiroshima-u.ac.jp}
\affiliation{Department of Quantum Matter, ADSE, Hiroshima University, Higashi-Hiroshima 739-8530, Japan}
\author{Takumi Hasegawa}
\affiliation{Department of Transdiciplinary Science and Engineering, ADSE, Hiroshima University, Higashi-Hiroshima 739-8521, Japan}
\author{Ryuma Nakajima}
\affiliation{Department of Quantum Matter, ADSE, Hiroshima University, Higashi-Hiroshima 739-8530, Japan}
\author{Kenshin Kurauchi}
\affiliation{Department of Quantum Matter, ADSE, Hiroshima University, Higashi-Hiroshima 739-8530, Japan}
\author{Satoshi Tsutsui}
\affiliation{Japan Synchrotron Radiation Research Institute (JASRI), SPring-8, Sayo, Hyogo 679-5198, Japan}
\affiliation{Institute of Quantum Beam Science, Graduate School of Science and Engineering, Ibaraki University, Hitachi 316-8511, Japan}
\author{Daisuke Ishikawa}
\affiliation{Materials Dynamics Laboratory, RIKEN SPring-8 Center, Sayo, Hyogo 679-5148, Japan}
\affiliation{Japan Synchrotron Radiation Research Institute (JASRI), SPring-8, Sayo, Hyogo 679-5198, Japan}
\author{Alfred Q. R. Baron}
\affiliation{Materials Dynamics Laboratory, RIKEN SPring-8 Center, Sayo, Hyogo 679-5148, Japan}
\affiliation{Japan Synchrotron Radiation Research Institute (JASRI), SPring-8, Sayo, Hyogo 679-5198, Japan}
\author{Hiroshi Tanida}
\affiliation{Liberal Arts and Science, Toyama Prefectural University, Imizu, Toyama 939-0398, Japan}


\date{\today}

\begin{abstract}
Soft phonon modes in tetragonal CeCoSi, which undergoes a structural transition at $T_0=12$ K followed by antiferromagnetic order at $T_{\text{N}}=9.5$ K, have been investigated using high-resolution inelastic x-ray scattering. Pronounced softening was detected in the transverse acoustic modes corresponding to the $(yz+zx)$-type monoclinic distortion, consistent with the experimentally determined triclinic structure. 
Remarkably, the softening persists up to the zone boundary along (0, 0, $q$), indicating a short correlation length of the lattice instability. 
This instability, characterized by a Curie-type strain susceptibility, is interpreted as a consequence of the on-site $4f$-$5d$ hybridization, which is intrinsic to this crystal structure due to the lack of inversion symmetry at the two Ce sites. 
\end{abstract}


\maketitle


\section{Introduction}
The breaking of space-inversion symmetry at magnetic ion sites has attracted considerable attention as a source of unconventional phenomena in magnetic materials.  
In particular, hybridization between orbitals of different parity can give rise to effects that are forbidden in centrosymmetric environments. 
In the superconductor CeRh$_2$As$_2$, which crystallizes in the CaBe$_2$Ge$_2$-type structure (space group $P4/nmm$), the emergence of multiple superconducting phases has been attributed to the sublattice degree of freedom and the absence of inversion symmetry at the two Ce sites in the unit cell~\cite{Khim21,Kibune22,Landaeta22,Ogata23,Semeniuk23}. 
Even in the normal state, the complex magnetic phase diagram suggests a quadrupolar contribution, arising from the involvement of high-energy crystalline electric field (CEF) excited states, which usually play only a minor role at low temperatures~\cite{Hafner22,Schmidt24}.  
In isostructural UPt$_2$Si$_2$, a strong interplay between the $5f$ electrons and the charge density wave (CDW) of $5d$ electrons in the Pt layer is attributed to the noncentrosymmetric environment of the U site~\cite{Kon24}. 

\begin{figure}[t]
\begin{center}
\includegraphics[width=7.5cm]{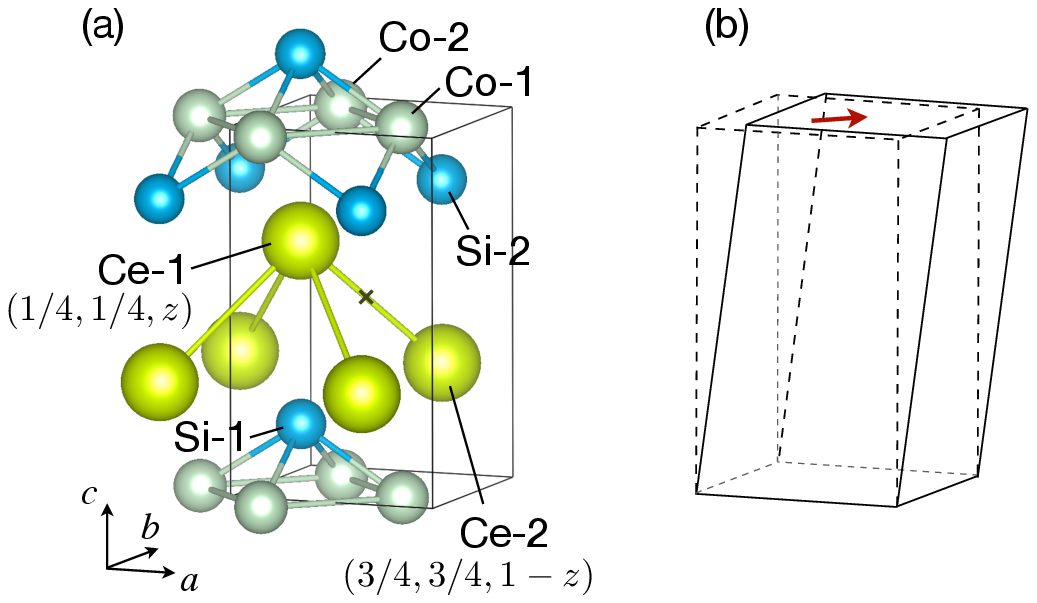}
\end{center}
\caption{(a) Crystal structure of CeCoSi drawn using VESTA~\cite{Momma11}; 
$a$=4.057 \AA, $c$=6.987 \AA, $z$(Ce)=0.678, Ce site symmetry $4mm$~\cite{Tanida19}. 
The cross at the center of the unit cell indicates one of the inversion centers. 
(b) Schematic illustration of the triclinic lattice distortion below $T_0=12$ K. 
}
\label{fig:Cryst}
\end{figure}

In this context, the nonmagnetic ordered phase of tetragonal CeCoSi, also belonging to $P4/nmm$, has attracted interest from a symmetry perspective. 
The crystal structure is shown in Fig.~\ref{fig:Cryst}(a). The two Ce ions in the unit cell are not at centrosymmetric sites but are interchanged by space inversion. CeCoSi undergoes a nonmagnetic transition at $T_0 =$12 K, followed by an antiferromagnetic transition at $T_{\text{N}}$=9.5 K~\cite{Tanida18,Tanida19,Manago21,Manago23}. 
X-ray diffraction revealed that the transition at $T_0$ is a structural transition to a triclinic phase~\cite{Matsumura22}. The $c$-axis tilts approximately in the [1 1 0] direction, as illustrated in Fig.~\ref{fig:Cryst}(b). 

The transition at $T_0$ exhibits several notable features. 
First, $T_0$ increases in magnetic fields, and the magnetic phase diagram of the ordered phase exhibits strong magnetic anisotropy, indicating that Ce-$4f$ is involved in the transition~\cite{Hidaka22,Hidaka25,Kanda24}. 
Consistently, this transition is absent in the non-$4f$ analog LaCoSi~\cite{Kawamura20}. 
Second, $T_0$ increases under high pressure, reaching $\sim 40$ K at 1.5 GPa, accompanied by a resistivity jump indicative of a CDW~\cite{Lengyel13}. 
It is noted that the Ce valence remains stable and trivalent up to $\sim 2$ GPa, while a separate structural transition around 4.9 GPa is likely associated with a Ce valence transition~\cite{Tanida18,Kawamura22}.

Although the triclinic transition may be regarded as a ferroquadrupole order of $O_{yz}+O_{zx}$, it is difficult to explain how the quadrupolar degrees of freedom arise in the well isolated $\Gamma_{7(1)}$ CEF ground doublet~\cite{Yatsushiro22}. 
The first and second excited doublets, $\Gamma_{7(2)}$ and $\Gamma_{6}$, are located at 120 K and 160 K, respectively, far above $T_0$, making it difficult to realize conventional quadrupole order~\cite{Nikitin20}. 
Nevertheless, the anisotropic magnetic phase diagram and the domain selection in magnetic fields can be phenomenologically described in terms of ferroquadrupole order~\cite{Matsumura22,Hidaka25,Ishitobi25}. Thus, the microscopic mechanism driving the structural transition remains elusive. 
Recent first-principles calculation to evaluate the multipolar Ruderman-Kittel-Kasuya-Yosida interactions predicted a high susceptibility toward $q=0$ CDW, suggesting a charge imbalance between Ce-1 and -2~\cite{Yamada24}. 
The optical conductivity measurement indicates a band modification near the Fermi level~\cite{Kimura23}. 
However, no direct experimental evidence has yet been obtained to clarify the origin of the structural transition. 

This study aims to elucidate the mechanism underlying the structural transition by investigating the lattice dynamics and soft phonon modes using high-resolution inelastic x-ray scattering (IXS). We report the observation of soft acoustic mode corresponding to the structural transition at $T_0$ and extending over a wide region of the Brillouin zone (BZ). 
Then, we propose that the on-site $4f$-$5d$ hybridization, enabled by the absence of inversion symmetry at the two Ce sites, plays a fundamental role for the lattice instability. 
We show that the symmetry lowering by the distortion leads to a decrease in the energy of the $4f$ ground state. 
Furthermore, the inter-site correlations mediated by $d$-$d$ transfer between Ce ions lead to a cooperative instability, giving rise to a Curie-law temperature ($T$) dependence of the softening toward the structural transition.
This work provides critical insight into how broken inversion symmetry couples to $f$-electron states and leads to lattice instability.

\section{Experiment}
We used the same single-crystalline sample previously used to study the structural transition~\cite{Matsumura22}. 
IXS experiment was performed using the high-resolution spectrometer at the RIKEN Quantum NanoDynamics beamline, BL43LXU, in SPring-8~\cite{Baron10,Baron16,Ishikawa21}. 
The spectrometer is equipped with an array of 7 (horizontal) $\times$ 4 (vertical) spherical analyzers using Si (11 11 11) reflection at 21.747 keV. 
The energy resolution was 1.4--1.5 meV depending on the analyzer. 
The measurement was performed using the flat $ab$-plane surface, where the scattering vector $\bm{Q}=\bm{k}'-\bm{k}$ was chosen near the (0 0 9) fundamental Bragg peak. 

\begin{figure}[t]
\begin{center}
\includegraphics[width=8.5cm]{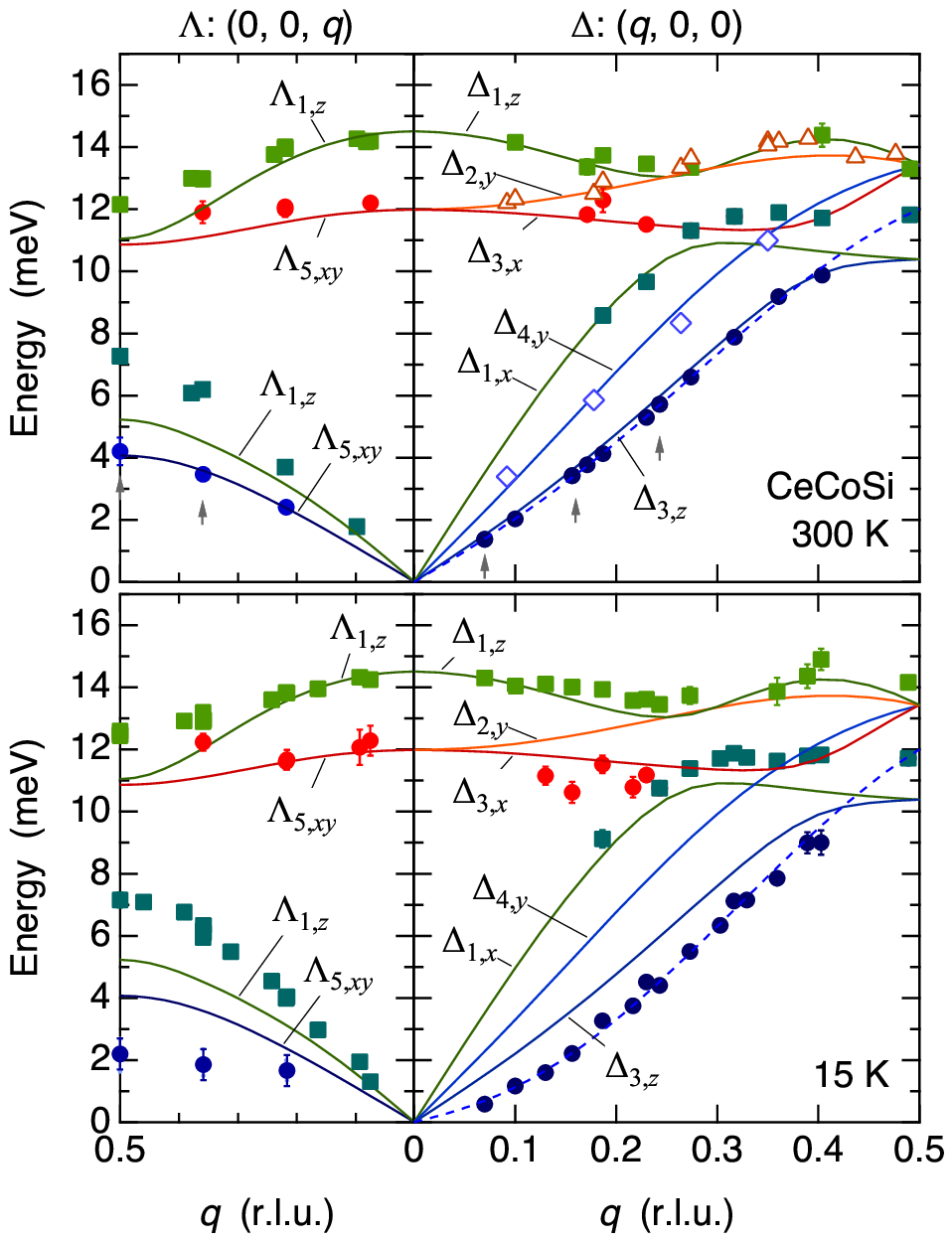}
\end{center}
\caption{Phonon-dispersion relations of Ce acoustic and optical branches along $\Delta$: $(q, 0, 0)$ and $\Lambda$: $(0, 0, q)$ at 300 K (top) and 15 K (bottom). 
The solid lines are the results of first-principles calculations, assuming that Ce-$4f$ is localized. The dashed lines are the fits to the data (see text). 
The subscript of the mode index denotes the irreducible representation and the direction of atomic displacement~\cite{SM}. 
Arrows indicate the data points whose $T$-dependences are shown in Fig.~\ref{fig:SpecH0L}(d,e).} 
\label{fig:Disp300K15K}
\end{figure}

\section{Results and Analysis}
Figure~\ref{fig:Disp300K15K} shows the phonon dispersion relations for the acoustic and optical branches of Ce along $(q, 0, 0)$ ($\Delta$-axis) and $(0, 0, q)$ ($\Lambda$-axis) in the first BZ. The closed symbols represent data obtained from scans along $\bm{Q}=(q, 0, 9)$ and $(0, 0, 9-q)$ using scattering from the $ab$-plane. 
The open symbols represent data obtained from scans along $\bm{Q}=(q, 5, 0)$ using the $ac$-plane side surface, which had some roughness.
These measurements for acoustic $\Delta_{4,y}$ and optical $\Delta_{2,y}$ branches were performed only at 300 K. 
The constant-$Q$ energy spectra were fitted using a damped harmonic oscillator function convoluted with the instrumental resolution. 
Mode assignments were carried out by comparing the $q$ dependence of the energy and intensity with the results of first-principles calculations~\cite{SM} (see also references \cite{Gonze20,Romero20,Blochl94,Torrent08,Perdew92,Monkhorst76,Pfrommer97,Gonze97a,Gonze97b,Holzwarth01} therein). 
The calculation, shown by the solid lines in Fig.~\ref{fig:Disp300K15K}, treats Ce-$4f$ as localized core electron, neglecting mixing with other orbitals of Ce, Co, and Si.

\begin{figure*}[t]
\begin{center}
\includegraphics[width=16cm]{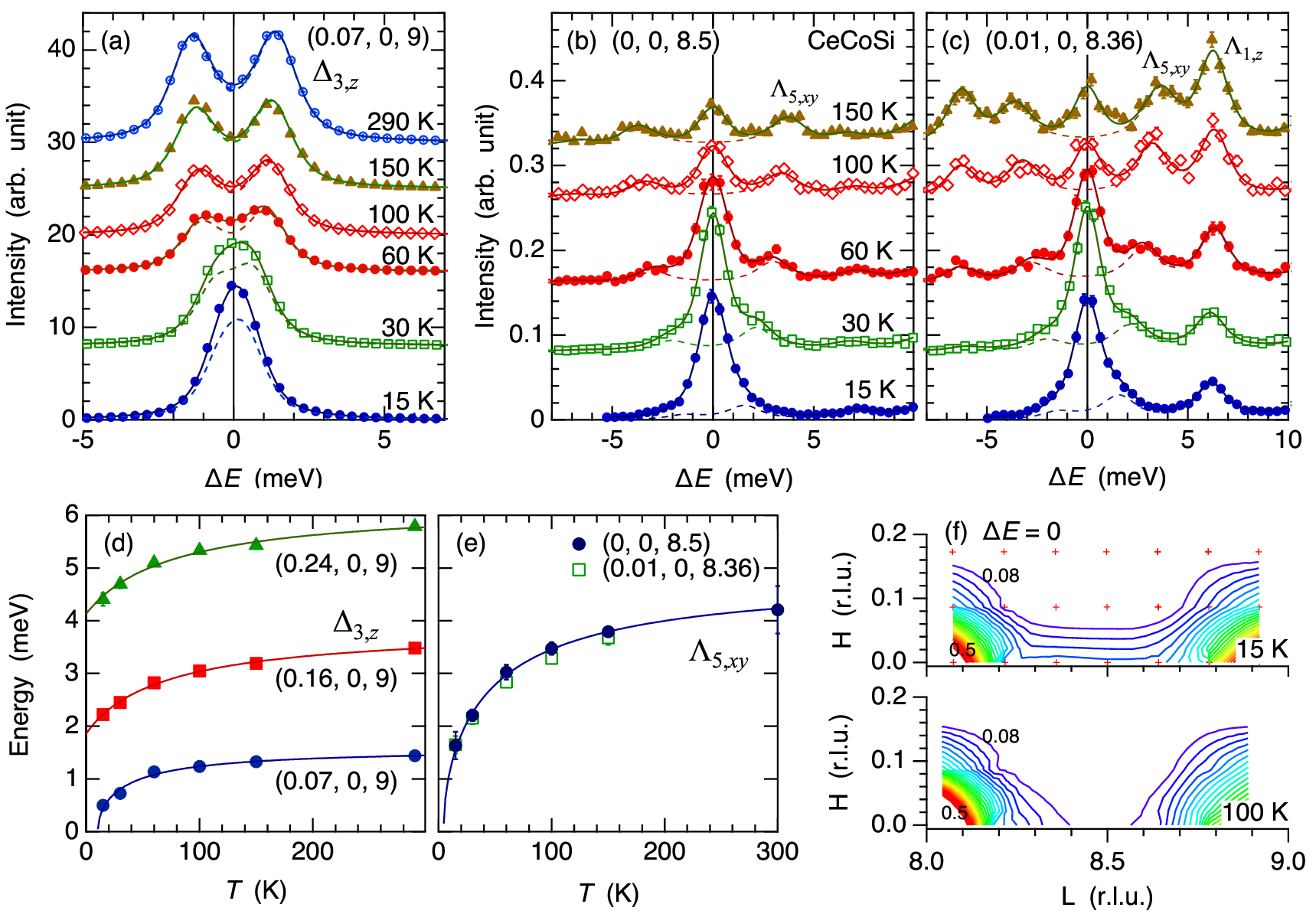}
\end{center}
\caption{(a,b,c) Temperature dependence of the phonon excitation spectra at $\bm{Q}=(0.07, 0, 9)$, $(0, 0, 8.5)$, $(0.01, 0, 8.36)$, respectively. 
The solid lines are the fits to the data. The dashed lines represent the inelastic contributions. 
(d,e) Temperature dependence of the excitation energies for the transverse acoustic modes $\Delta_{3,z}$ and $\Lambda_{5,xy}$, respectively. The solid lines are fits assuming a Curie-Weiss temperature dependence of the strain susceptibility. 
(f) Intensity contour map of the elastic ($\Delta E=0$) component at 15 K and 100 K.  Contour levels range from 0.08 to 0.5 in equal steps. The $+$ symbols indicate the centers of the analyzers. 
}
\label{fig:SpecH0L}
\end{figure*}

At 300 K, the observed dispersion relations are well reproduced by the first-principles calculation. 
The localized nature of Ce-$4f$ is consistent with the known bulk properties at ambient pressure~\cite{Tanida19,Hidaka22,Kanda24}.
The concave upward relation of the $\Delta_{3,z}$ transverse acoustic (TA) mode is explained well. 
Although some discrepancies remain in the absolute energy scale for the $\Lambda_{1,z}$ longitudinal acoustic (LA) mode, the $\Lambda_{1,z}$ and $\Lambda_{5,xy}$ optical modes at $q=0.5$, and the $\Delta_{3,z}$-$\Delta_{1,x}$ degeneracy at $q=0.5$, they are not essential for the following discussions as they are likely due to the subtle treatment of the $4f$ electron.

The bottom panels of Fig.~\ref{fig:Disp300K15K} show the data at 15 K, revealing a clear softening of the $\Lambda_{5,xy}$ and $\Delta_{3,z}$ TA modes. The $\Sigma_{4,z}$ TA mode, propagating along the $[1 1 0]$ axis, also exhibits similar softening~\cite{SM}. 
Notably, the $\Lambda_{5,xy}$ mode exhibits softening throughout the BZ, indicating a short correlation length of the lattice instability along the $c$-axis, estimated to be only 1--2 unit cells. 
The dashed line for the $\Delta_{3,z}$ mode is a fit to estimate the correlation length $\xi$, using renormalized frequency expressed by 
$\omega(q)=\sqrt{C_0/\rho/\{1+\lambda^2\chi_0/C_0(1+\xi^2 q^2)\}}\sin \pi q$~\cite{Fernandes10,Weber18,Merritt20}, where $\chi_0$ represents a local strain susceptibility for the dimer formation in the present case as discussed later. Along the $a$ axis, $\xi$ was estimated to be $\sim$5 unit cells, corresponding to the change in the initial slope of $\omega(q)$ around $q\sim 0.2$ r.l.u. 
The dashed curve at 300 K was assumed to represent the unsoftened dispersion. 
Except for the TA modes, the energies of the LA and optical modes do not exhibit any significant $T$-dependence.

The $T$-dependence of the energy spectrum shown in Fig.~\ref{fig:SpecH0L}(a) clearly demonstrates the elastic softening of the $\Delta_{3,z}$ TA mode, as observed at $\bm{Q}=(0.07, 0, 9)$ near the $\Gamma$ point. 
Figure~\ref{fig:SpecH0L}(b) and (c) shows the softening of the TA mode along the $\Lambda$ axis, observed at $\bm{Q}=(0, 0, 8.5)$ at the zone boundary and at $\bm{Q}=(0, 0, 8.36)$ within the BZ. The LA mode of $\Lambda_{1,z}$ remains constant at 6 meV regardless of temperature, showing no softening, whereas the $\Lambda_{5,xy}$ TA mode exhibits softening as the temperature decreases. 
In the geometry along (0, 0, $L$), the intensity of the TA phonon is expected to vanish. The observed intensity results from the finite $Q$ resolution of the analyzer. 
Although the fitting becomes difficult at 15 K due to limited resolution, the intensity variation is well described by the Bose factor. The oscillator strength remains constant and the line-width is close to the resolution limit. 
These results indicate relatively weak energy dissipation from the dimerization to other electronic channels, in contrast to the linewidth broadening typical of CDW transitions with strong electron-phonon coupling~\cite{Requardt02,Weber11,Leroux12,Blackburn13,Tacon13}.


The $T$-dependence of the TA-phonon energies is summarized in Figs.~\ref{fig:SpecH0L}(d,e). 
The softening begins at room temperature and follows a Curie-Weiss law, indicating that the shear strain reduces the free energy. 
The $T$-dependence was modeled using a simple Curie-Weiss form of the dimer strain susceptibility, assuming $\chi_q(T)=\beta/(T-\theta_q)$, where $\beta$ and $\theta_q$ are constants. 
For $\bm{Q}=(0.07, 0, 9)$ near the $\Gamma$ point, the phonon energy extrapolates to zero at 10.3 K, in reasonable agreement with $T_0$=12 K at $q$=0. At finite $q$, however, $\theta_q$ deviates from $T_0$ and takes a smaller value, and the phonon energy is expected to exhibit a cusp at $T_0$. 

Another important feature is the emergence of an elastic peak at low temperatures as clearly observed in Figs.~\ref{fig:SpecH0L}(a--c). 
This indicates the development of slow relaxation coexisting with the normal phonon oscillations. This is also related to the short correlation length along the $c$ axis, as demonstrated in Fig.~\ref{fig:SpecH0L}(f) by the contour map of the $\Delta E=0$ intensity extending over the entire BZ. 
However, we note that such elastic scattering is commonly observed in materials undergoing structural transitions, and its origin is sometimes debated in terms of defect-induced phenomenon~\cite{Tacon13,Shirane71,Shapiro72,Cowley06,Hoesch09,Ohwada18}.

\section{Discussion}
Although the elastic softening can be phenomenologically explained by introducing a conventional coupling between $4f$ quadrupole and shear strains, this approach presents a difficulty due to the need to incorporate the CEF excited state as high as 120 K. 
Here, we focus instead on the noncentrosymmetric environment of the Ce ions, which allows the bilinear coupling between strain and atomic displacement. This odd-parity term takes the form 
\begin{equation}
E_{\text{odd}}=\sum_{i,\alpha}\lambda_{i,\alpha}e_{i}u_{\alpha} \;,
\end{equation}
where $e_{i}$ $(i=1\sim6)$ denotes the strain component, $u_{\alpha}$ ($\alpha=x,y,z$) the shift of the equilibrium atomic position, and $\lambda_{i,\alpha}$ the coupling constant.  
In fact, one of the motivations for performing the IXS experiment was the prediction of soft modes by the first-principles calculations. 
When Ce-$4f$ is allowed to hybridize with other orbitals, the TA modes of $\Delta_{3,z}$, $\Lambda_{5,xy}$, and $\Sigma_{4,z}$ become unstable, exhibiting negative energies near the $\Gamma$ point~\cite{SM}. This indicates that CeCoSi has an intrinsic instability toward monoclinic distortion driven by the Ce-$4f$ hybridization.  

\begin{figure}[t]
\begin{center}
\includegraphics[width=8.5cm]{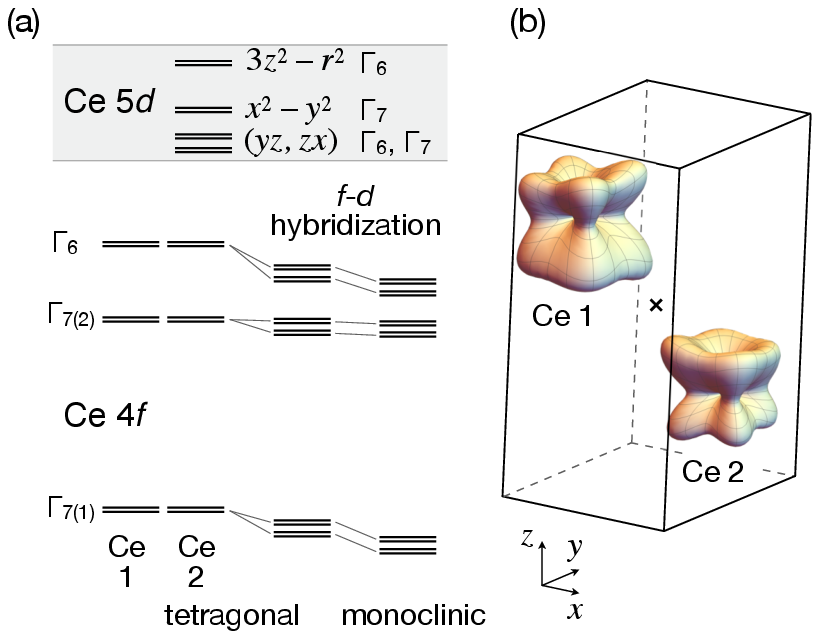}
\end{center}
\caption{(a) Schematic illustration of the energy levels including local $4f$-$5d$ hybridization and monoclinic distortion. 
The splitting of the $f$ levels in the hybridized state is exaggerated for clarity.
(b) Calculated charge distribution of the $4f$-$5d$ hybridized ground state involving Ce-1 and Ce-2. 
}
\label{fig:dfmixing}
\end{figure}

The microscopic origin of the bilinear coupling is most likely the on-site $4f$-$5d$ hybridization enabled by the finite local electric field arising from the absence of inversion symmetry at the Ce sites. 
Once the $4f$-$5d$ hybridization is included, the energies of the three CEF Kramers-doublets are lowered through second-order perturbation effects. 
Furthermore, intersite $d$-$d$ transfer between Ce ions couples the two $\Gamma_{7(1)}$ doublets of Ce-1 and Ce-2, leading to a dimerized state, as schematically illustrated in Fig.~\ref{fig:dfmixing}(a).  
The energy lowering is driven by the mixing with $5d$-$\Gamma_7$ states such as $|yz\!\uparrow\rangle - i|zx\! \uparrow\rangle$ and $|yz\! \downarrow\rangle + i|zx\! \downarrow\rangle$. 
A more detailed qualitative explanation is provided in the Supplemental Material~\cite{SM}. 
A similar discussion can also be found in \cite{Moulding25}. 

When a lattice distortion is introduced, the $5d$-$\Gamma_7$ and $5d$-$\Gamma_6$ states mix with each other, giving rise to a finite matrix element $\langle 4f\text{-}\Gamma_{7(1)}| \mathcal{H}_{fd} | 5d\text{-}\Gamma_6 \rangle$. 
In addition, Ce displacements within the $xy$-plane tilt the $E_z$ field and generate $E_x$ and $E_y$, which further enhance the matrix element. 
As a result, the energy levels of the Ce-$4f$ CEF states are further lowered, as illustrated in the monoclinic level scheme in Fig.~\ref{fig:dfmixing}(a). 
Using the $5d$-CEF and $f$-$d$ hybridization parameters of \cite{Yamada24} (namely, a total $5d$-CEF splitting of $\sim$1.4 eV and an $f$-$d$ hybridization of $\sim$ 100 meV), and by setting the bare $d$ level at 500 meV above the $f$ level, an energy gain of $\sim$1 meV -- sufficient to account for $T_0$ -- is obtained by assuming a $5d$-CEF tilt of $\sim1.5^{\circ}$ and an $E_z$-field tilt of $\sim10^{\circ}$~\cite{SM}. 
This is consistent with the relationship that the transverse displacements $u_x$ and $u_y$ couple directly to $e_5$ ($e_{zx}$) and $e_4$ ($e_{yz}$), respectively. 
Although the structure ultimately relaxes to triclinic to minimize the energy, the essential instability originates from the monoclinic distortion. 
In contrast, distortions of $xy$-, $(x^2-y^2)$-, or $(3z^2-r^2)$-type do not provide an energy gain because they do not tilt the $E_z$-field.  

Although one might question how such a small energy scale can result from a large hybridization energy and a substantial energy-level structure, this is actually not surprising by analogy with the magnetic exchange interactions, which are also on the order of meV and arise from $c$-$f$ hybridization between conduction and $f$ electrons involving energy scales on the order of eV.  The $c$-$f$ exchange is typically expressed as $J_{cf} = V^2/U$, where the hybridization $V$ and the local Coulomb repulsion $U$ are both on the order of eV. 

The charge distribution of the ground doublet under monoclinic distortion is illustrated in Fig.~\ref{fig:dfmixing}(b), with the $4f$-$5d$ hybridization taken into account.
It retains the main features of the original $4f$-$\Gamma_{7(1)}$ distribution, with an additional $5d$ component linking Ce-1 and Ce-2. 
Evaluating the electric quadrupoles, we see that all five components are identical at Ce-1 and Ce-2, indicating a ferroquadrupole ordering. Importantly, this ordering arises without requiring quadrupolar degrees of freedom in the original CEF ground state.

The bonding state between Ce-1 and Ce-2 can be regarded as a dimerization mediated by the $5d$ electrons.
This dimerization is realized by selecting one of the four nearest-neighbor Ce-Ce bonds. 
The associated degree of freedom gives rise to the strain susceptibility and naturally leads to Curie-type elastic softening, which can be derived straightforwardly~\cite{SM}. 
While such softening usually requires orbital degeneracy within a single-ion framework, in the present case it originates from the coupling of two Kramers doublets through $4f$-$5d$ hybridization. 

As evidenced by the pronounced softening of the $\Lambda_{5,xy}$ mode throughout the BZ, also by the diffuse scattering shown in Fig.~\ref{fig:SpecH0L}(f), the correlation length of the distortion is extremely short along the $c$-axis at 15 K. 
Along the $a$-axis, in contrast, the correlation length is longer, and the local dimer formation extends over about five unit cells.  
This anisotropy in the dynamical correlation likely reflects the difference in $d$-$d$ transfer between Ce ions: along the $c$-axis the transfer is interrupted by the Co and Si layers, whereas in the $ab$-plane it proceeds directly within the Ce layer. 

The present $f$-$d$ hybridization formalism can in principle be extended to hybridization with Co-$3d$ states. 
However, the observed softening of the TA phonon modes shows that the transverse displacement of the heavy Ce atoms couples to the strain. 
In contrast, no softening is seen in the Co optical modes, indicating that Co displacements do not lead to an energy gain. 
Moreover, the Co-$3d$ hybridization is of inter-atomic transfer type, and the small changes in atomic distance have only a minor effect on the matrix element and the associated energy lowering. 
Such hybridization typically leads instead to the Kondo effect, reflected in the bulk modulus softening, and does not couple to shear strain~\cite{Bruls90,Nemoto96,Shimizu96}. 


The pronounced lattice instability in CeCoSi may be associated with the characteristic that the $\Gamma_{7(1)}$ wavefunction extends toward neighboring Ce ions, thereby making the dimerization mechanism via $d$-$d$ transfer more effective. 
By contrast, in CeRh$_2$As$_2$ and UPt$_2$Si$_2$, the square-lattice layers of magnetic ions are separated by other atoms, leading to different symmetry conditions.
Nevertheless, since $f$-$d$ hybridization can generally induce lattice instabilities at locally noncentrosymmetric sites, it may also play an important role in these compounds.
A related case is CePd$_2$Al$_2$, which has the CaBe$_2$Ge$_2$-type structure and exhibits a CeCoSi-like transition that is even more pronounced~\cite{Klicpera15,Klicpera17}. 

\section{Conclusion}
We observed soft acoustic phonon modes in CeCoSi using high-resolution IXS, uncovering the lattice instability responsible for the structural transition at $T_0=12$ K. 
All soft modes are associated with the $(yz+zx)$-type monoclinic distortion. Notably, the softening of the TA mode along $(0, 0, q)$ extends to the zone boundary, indicating an extremely short correlation length. This instability, characterized by a Curie-type strain susceptibility, is naturally explained by the intrinsic on-site $4f$-$5d$ hybridization enabled by the absence of inversion symmetry at the Ce sites, which gives rise to bilinear coupling between the $yz$- and $zx$-shear strains and the Ce atomic displacements within the $xy$-plane.


\begin{acknowledgments}
The authors acknowledge valuable discussions with T. Aoyama and T. Yamada.  
This work was supported by the JSPS Grant-in-Aid for Scientific Research (B) (No. JP20H01854) and JSPS Grant-in-Aid for Transformative Research Areas (Asymmetric Quantum Matters, No. JP23H04867). 
The synchrotron radiation experiments were performed at the BL43LXU of SPring-8 with the approval of the Japan Synchrotron Radiation Research Institute (JASRI) (Proposal No. 2023A1310).

\end{acknowledgments}

\bibliography{CeCoSi_IXS}

\clearpage
\begin{widetext}

\begin{center}
\textbf{\Large{Supplemental Material}}
\end{center}
\vspace{2mm}

\begin{center}
\textbf{\large{Acoustic phonon softening and lattice instability driven by on-site $f$-$d$ hybridization in CeCoSi}} \\
\vspace{4mm}
T. Matsumura, T. Hasegawa, R. Nakajima, K. Kurauchi, S. Tsutsui, D. Ishikawa, A. Q. R. Baron, and H. Tanida
\end{center}
\vspace{5mm}

\renewcommand{\topfraction}{1.0}
\renewcommand{\bottomfraction}{1.0}
\renewcommand{\dbltopfraction}{1.0}
\renewcommand{\textfraction}{0.01}
\renewcommand{\floatpagefraction}{1.0}
\renewcommand{\dblfloatpagefraction}{1.0}
\setcounter{topnumber}{5}
\setcounter{bottomnumber}{5}
\setcounter{totalnumber}{10}

\renewcommand{\theequation}{S\arabic{equation}}
\renewcommand{\thefigure}{S\arabic{figure}}
\renewcommand{\thetable}{S-\Roman{table}}
\setcounter{section}{0}
\setcounter{equation}{0}
\setcounter{figure}{0}
\setcounter{page}{1}

\begin{center}
\textbf{\large{I. Irreducible representation of the phonon modes}}
\end{center}

The irreducible representation of the phonon mode at the wave vector $\bm{k}$ is expressed as
\begin{align}
\Psi_{\eta} (\bm{k}) &= \sum_{n} \sum_{i} \exp[i\bm{k}\cdot(\bm{t}_n+\bm{d}_i)] \Phi_{\eta,i}(\bm{k}) \;, \\
\Phi_{\eta,i}(\bm{k}) &= (\alpha_{\eta,i} \hat{\bm{x}} + \beta_{\eta,i} \hat{\bm{y}} + \gamma_{\eta,i} \hat{\bm{z}}) \;,
\end{align}
where $\eta$ labels the irreducible representation, and $\bm{t}_n + \bm{d}_i$ denotes the position of the $i$-th atom in the $n$-th unit cell. 
When considering only the Ce atoms, $i$ runs over 1 and 2. 
$\Phi_{\eta,i}(\bm{k})$ describes the displacement component of the $i$-th atom associated with the irreducible representation $\eta$ at the wave vector $\bm{k}$. $\alpha$, $\beta$, and $\gamma$ are the coefficients of the displacement vector. 
The six phonon modes of the Ce atoms are illustrated in Fig.~\ref{fig:PhononModes}. These labels are used throughout the main text. 
It is noted that the phase factor $\exp(i\bm{k}\cdot\bm{d}_i)$ is outside the expression of $\Phi_{\eta,i}(\bm{k})$.

\begin{figure*}[b]
\begin{center}
\includegraphics[width=12cm]{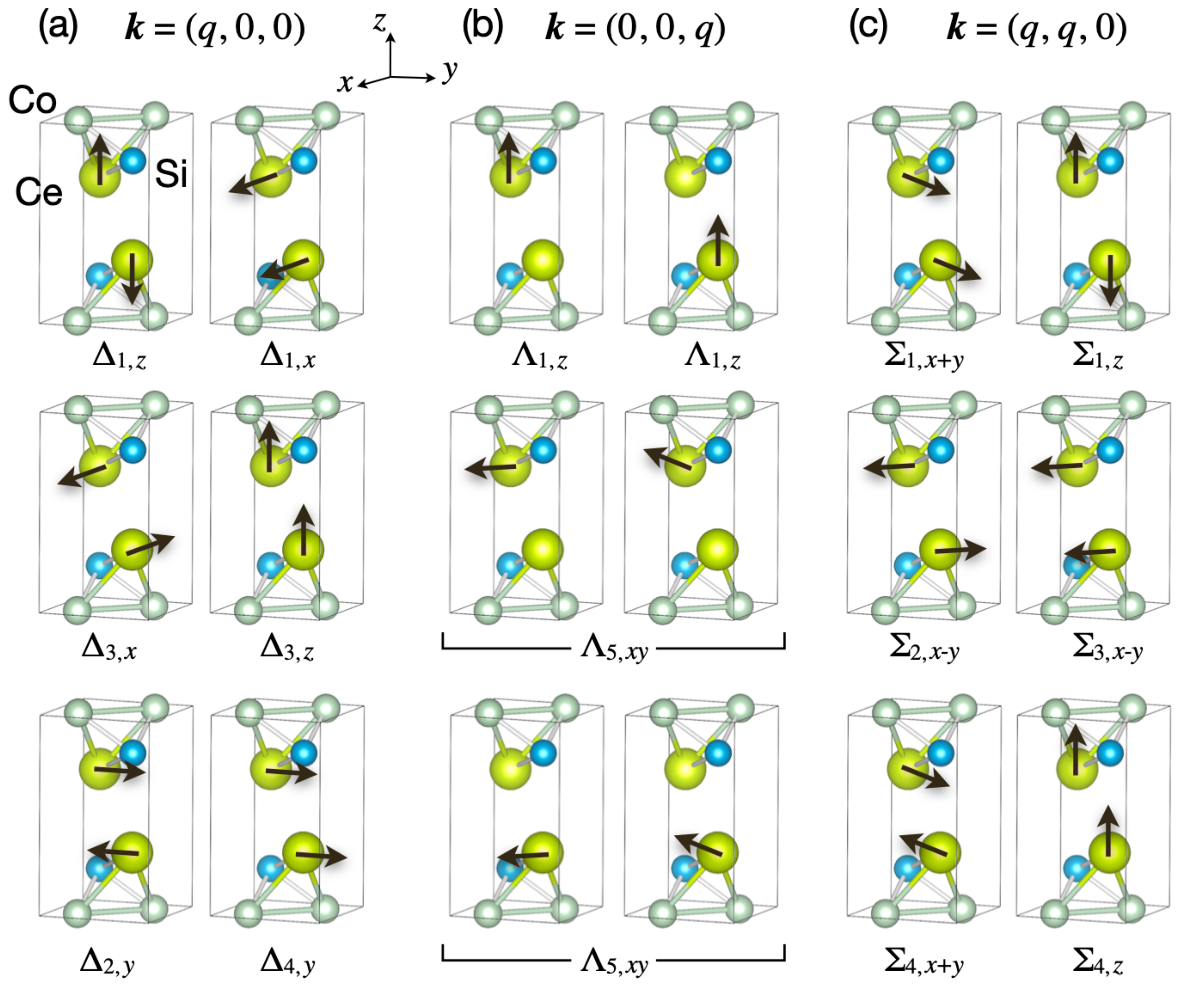}
\end{center}
\caption{Six phonon modes of the Ce atoms for wave vectors along (a) $\Delta: (q, 0, 0)$, (b) $\Lambda: (0, 0, q)$, and (c) $\Sigma: (q, q, 0)$. 
The subscript in each mode index represents the irreducible representation and the direction of atomic displacement.  }
\label{fig:PhononModes}
\end{figure*}

\newpage
\begin{center}
\textbf{\large{II. Phonon dispersion relation at 300 K and 15 K}}
\end{center}

Figure~\ref{fig:DispAll300K15K} displays the complete set of phonon dispersion relations obtained in this work at 300 K and 15 K. 
The optical branches in the 12--15 meV range originate from the Ce atoms, and those in the 18--26 meV range are from the Co atoms. 
The optical branches of the Si atoms are in the 40--50 meV range. 
The $\Delta_{4,y}$ and $\Delta_{2,y}$ branches associated with the Ce atoms were measured using the $ac$-plane surface. 
Since the sample was a thin platelet with a flat $ab$-plane surface, the $ac$-plane corresponded to the side edge of the sample. The measurement was relatively challenging due to the surface roughness. Only the peaks corresponding to these two modes could be identified in the scans along $\bm{Q}=(q, 5, 0)$. 
The measurements of the $\Delta_{4,y}$ and $\Delta_{2,y}$ branches were not performed at 15 K. 

\begin{figure*}[b]
\begin{flushleft}
\hspace{5mm}\includegraphics[width=17.2cm]{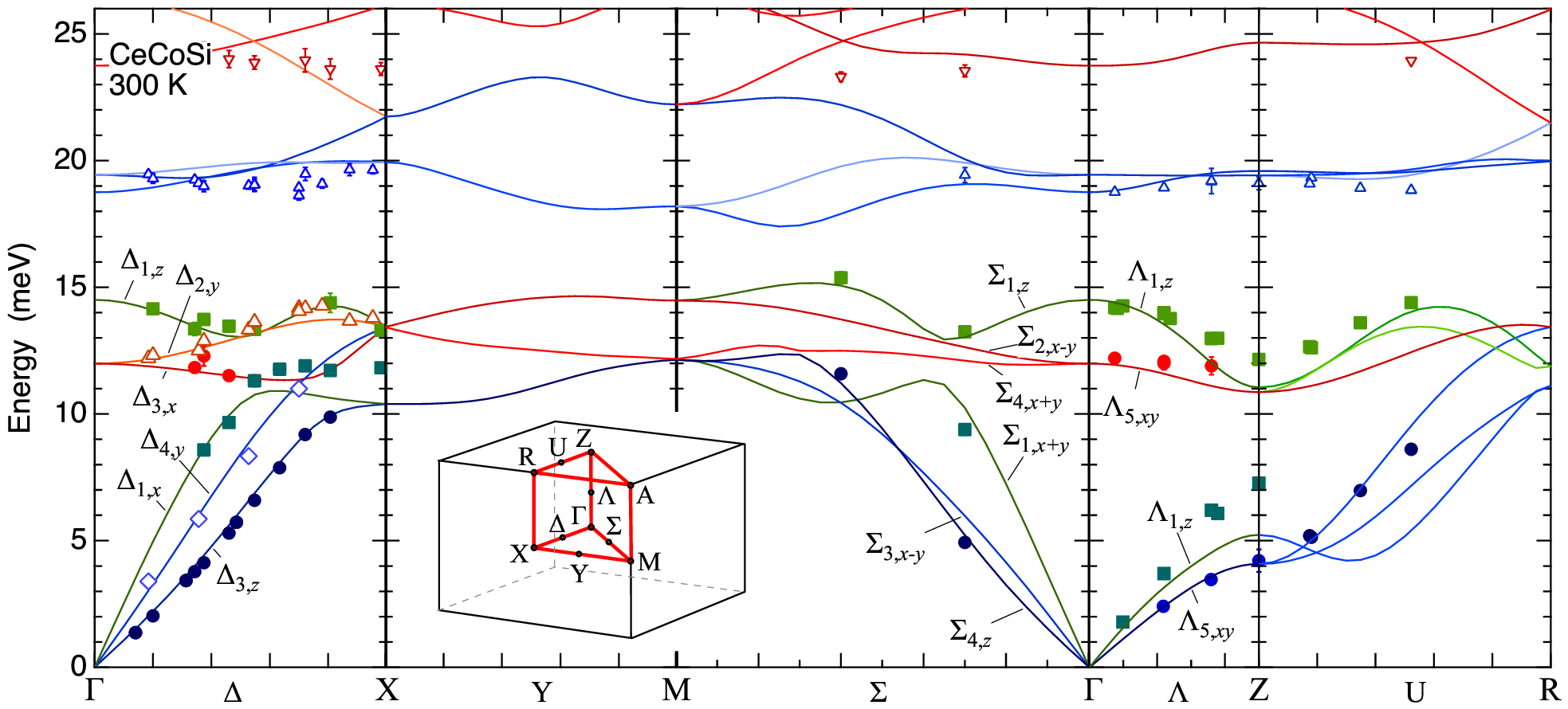}\\
\end{flushleft}
\begin{flushleft}
\hspace{5mm}\includegraphics[width=14cm]{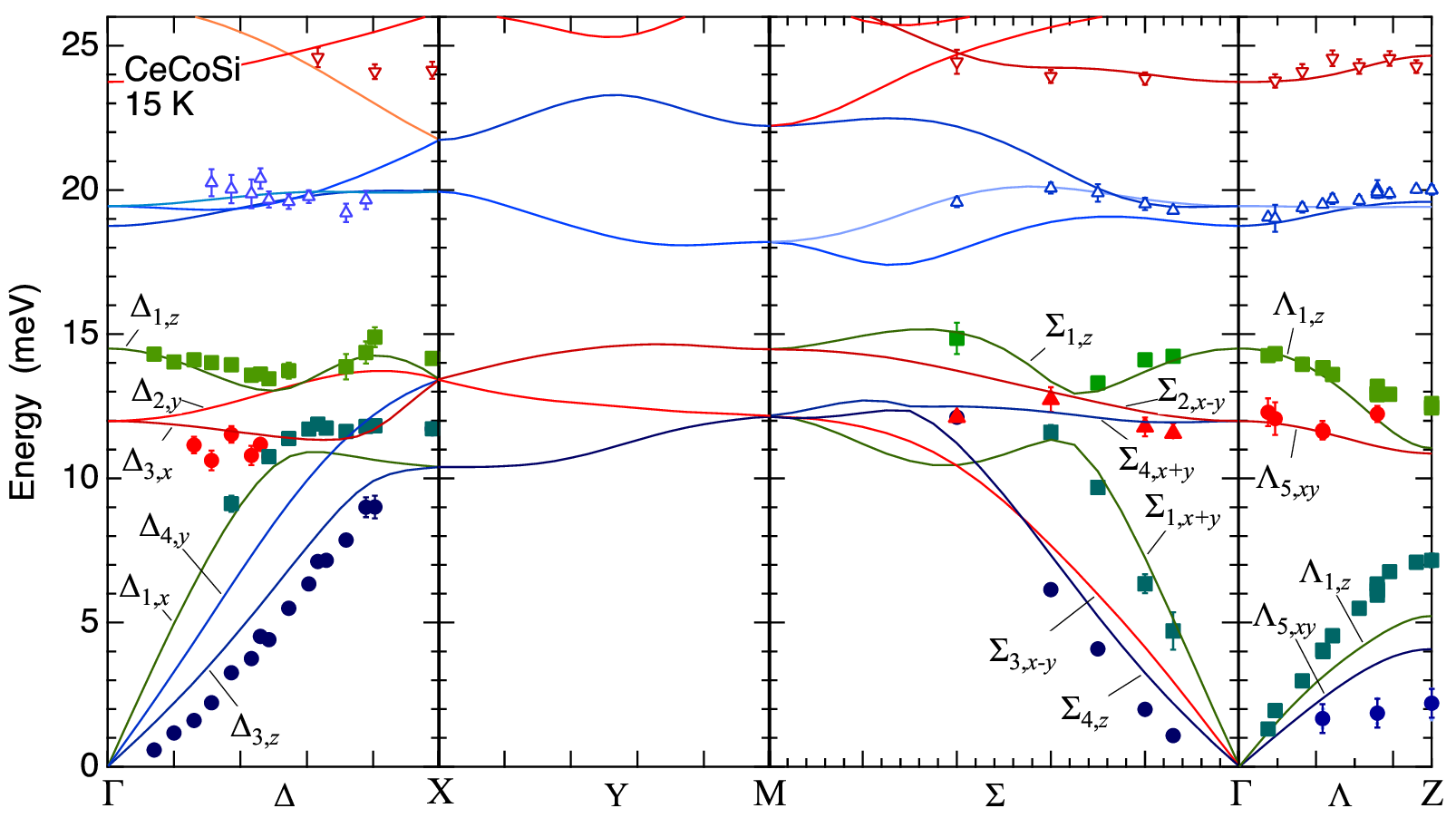}
\end{flushleft}
\caption{All the phonon dispersion relations measured in this work at 300 K and 15 K. The solid lines represent the results of first-principles calculations. The inset shows the first Brillouin zone of CeCoSi. }
\label{fig:DispAll300K15K}
\end{figure*}

\newpage
\begin{center}
\textbf{\large{III. First-principles calculation method and phonon dispersions allowing Ce-$4f$ hybridization}}
\end{center}

First-principles calculations were performed using the ABINIT package [1, 2]. 
We employed the projector augmented wave (PAW) method [3, 4] with a plane-wave cutoff energy of 381 eV, local density approximation (LDA) with the exchange-correlation potential by Perdew and Wang [5], $12 \times 12 \times 8$ Monkhost-Pack $k$-point mesh [6], and Fermi-Dirac smearing at 158 K. 
The crystal structure was optimized using the Broyden-Fletcher-Goldfarb-Shanno algorithm [7], ensuring that the residual forces were reduced to below 0.0026 eV/\AA. 
Phonon dispersions were obtained from the dynamical matrices calculated on $6 \times 6 \times 4$ $k$-point mesh using density functional perturbation theory (DFPT) [8, 9]. 
The PAW atomic datasets were generated using ATOMPAW [10]. The valence configurations were set to $4s^{2}4p^{0}3d^{7 }$ for Co and $3s^{2}3p^{2}$ for Si, with an additional projector for each angular momentum $l$ up to $l=2$. 
The augmentation radii $r_{\text c}$ were set to 1.22 \AA\ for Ce, and 1.11 \AA\ for both Co and Si. 

Two different electronic configurations for Ce were examined. 
In the ``frozen-$f$" configuration, the $4f^{1}$ electron was treated as part of the core, resulting in a spherical $4f$ charge distribution and effectively eliminating the $4f$-orbital degrees of freedom.  
The valence configuration is $5s^{2}5p^{6}6s^{2}6p^{0}5d^{1}$, with an additional projector for $l=0$, 1, 2, and 3. 
The calculated phonon dispersion with this ``frozen-$f$" configuration is shown in Fig. 2 in the main text and in Fig. S2. 

In contrast, the ``free-$f$" configuration treats the $4f$ electron as part of the valence shell, with a configuration $5s^{2}5p^{6}6s^{2}6p^{0}5d^{1}4f^{1}$, and again including an additional projector for each $l$. The resulting dispersion is presented in Fig. S3. 
In this case, the $f$-electron becomes itinerant, leading to a reduced $f$-electron occupancy and a notable lowering of the optical phonon energies compared to both the experiment and the ``frozen-$f$" calculation. 

Importantly, in the ``free-$f$" configuration, the transverse acoustic modes $\Delta_{3,z}$, $\Lambda_{5,xy}$, and $\Sigma_{4,z}$ exhibit negative energies near the zone center, indicating dynamical instabilities. 
In particular, the $\Lambda_{5,xy}$ mode remains unstable across the entire Brillouin zone. These soft modes are consistent with experimental observations, whereas no such instability appears in the ``frozen-$f$" case. 
These results highlight the critical role of $f$-$d$ hybridization in driving the lattice instability.

\begin{figure*}[h]
\begin{center}
\includegraphics[width=15cm]{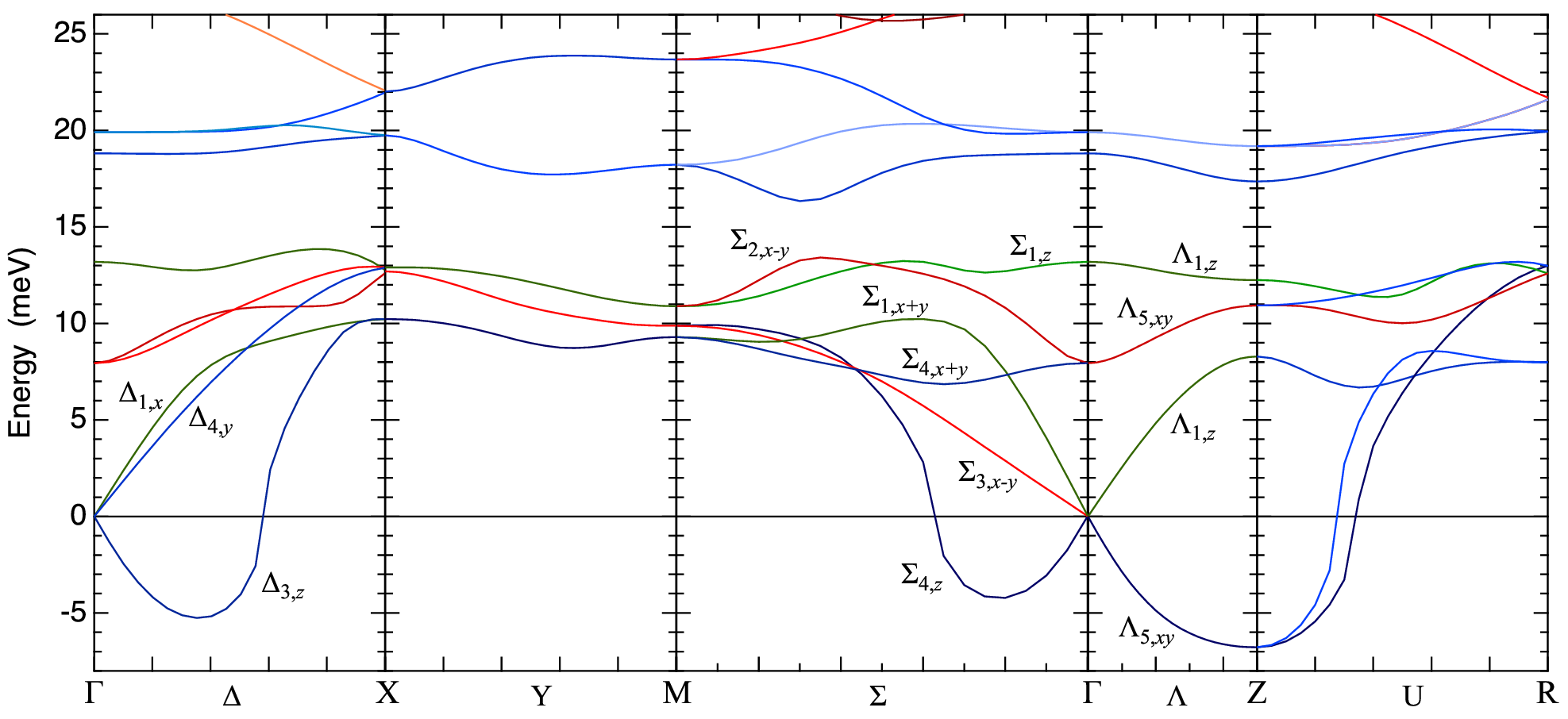}
\end{center}
\vspace{5mm}
\caption{First-principles calculation of phonon dispersion relation allowing Ce-$4f$ to hybridize with other orbitals.  }
\label{fig:DispItinerant}
\end{figure*}

\newpage
\begin{center}
\textbf{\large{IV. Lattice instability driven by on-site $4f$-$5d$ hybridization }}
\end{center}

This section explains how the $4f$-state energy is lowered by a lattice distortion through $f$-$d$ hybridization, providing a qualitative description of Fig.~\ref{fig:dfmixing} in the main text. 
This microscopic mechanism can be phenomenologically expressed in terms of the odd-parity bilinear term $\sum_{i,\alpha}\lambda_{i,\alpha}e_{i}u_{\alpha}$ ($i=1\sim 6$, $\alpha=x,y,z$), whose formalism is also presented here. 
This framework well accounts for the Curie-type elastic softening. 
A more detailed and quantitative analysis will be presented in a separate paper based on first-principles calculations. 

\vspace{5mm}
\begin{center}
\textbf{A. $5d$ orbital with spin-orbit coupling}
\end{center}

The $5d$ orbitals of Ce are expressed in the $|l=2, m\rangle$ ($m=2,1,\cdots,-2$) basis as
\begin{align}
|d_{yz}\rangle &= \frac{i}{\sqrt{2}}(|2,1\rangle + |2,\bar{1}\rangle) \;, \nonumber \\
|d_{zx}\rangle &= \frac{1}{\sqrt{2}}(-|2,1\rangle + |2,\bar{1}\rangle)  \;, \nonumber \\
|d_{xy}\rangle &= \frac{i}{\sqrt{2}}(-|2,2\rangle + |2,\bar{2}\rangle)   \label{eqS:dcef} \;,  \\
|d_{u}\rangle &= |2,0\rangle  \;,  \nonumber \\
|d_{v}\rangle &= \frac{1}{\sqrt{2}}(|2,2\rangle + |2,\bar{2}\rangle)  \;. \nonumber 
\end{align}
In the tetragonal crystalline-electric-field (CEF), the energy levels split into four levels, where $d_{yz}$ and $d_{zx}$ remains degenerate. 

When the spin-orbit coupling is introduced, the $5d$ orbitals are expressed as
\begin{align}
|d_{1a}\rangle &= c_{1yz}(i |d_{yz,\uparrow}\rangle - |d_{zx,\uparrow}\rangle) + c_{1u}( |d_{u,\downarrow}\rangle  \;, \nonumber \\
|d_{1b}\rangle &= c_{1yz}(i |d_{yz,\downarrow}\rangle + |d_{zx,\downarrow}\rangle) + c_{1u}(  |d_{u,\uparrow}\rangle  \;,  \nonumber \\
|d_{2a}\rangle &= c_{2yz}(i |d_{yz,\downarrow}\rangle - |d_{zx,\downarrow}\rangle) +c_{2v} |d_{v,\uparrow}\rangle - c_{2xy}|d_{xy,\uparrow}\rangle   \;, \nonumber \\
|d_{2b}\rangle &= c_{2yz} (i |d_{yz,\uparrow}\rangle + |d_{zx,\uparrow}\rangle) +c_{2v} |d_{v,\downarrow}\rangle + c_{2xy} |d_{xy,\downarrow}\rangle   \;, \nonumber \\
|d_{3a}\rangle &= c_{3yz} (i |d_{yz,\downarrow}\rangle - |d_{zx,\downarrow}\rangle) -c_{3v} |d_{v,\uparrow}\rangle  + c_{3xy} |d_{xy,\uparrow}\rangle   \;,  \label{eqS:dcefSO}\\
|d_{3b}\rangle &=  c_{3yz} (i |d_{yz,\uparrow}\rangle + |d_{zx,\uparrow}\rangle) - c_{3v} |d_{v,\downarrow}\rangle  -c_{3xy} |d_{xy,\downarrow}\rangle   \;, \nonumber \\
|d_{4a}\rangle &=  c_{4yz} (i |d_{yz,\uparrow}\rangle - |d_{zx,\uparrow}\rangle) - c_{4u} |d_{u,\downarrow}\rangle   \;,  \nonumber \\
|d_{4b}\rangle &= c_{4yz} (i |d_{yz,\downarrow}\rangle + |d_{zx,\downarrow}\rangle) - c_{4u} |d_{u,\uparrow}\rangle  \;.  \nonumber \\
|d_{5a}\rangle &=  c_{5yz} (i |d_{yz,\downarrow}\rangle - |d_{zx,\downarrow}\rangle) + c_{5v} |d_{v,\uparrow}\rangle + c_{5xy} |d_{xy,\uparrow}\rangle  \;, \nonumber \\
|d_{5b}\rangle &= c_{5yz} (i |d_{yz,\uparrow}\rangle + |d_{zx,\uparrow}\rangle) +  c_{5v} |d_{v,\downarrow}\rangle -  c_{5xy} |d_{xy,\downarrow}\rangle   \;, \nonumber 
\end{align}
The orbital degeneracy between $d_{yz}$ and $d_{zx}$ is lifted, resulting in five Kramers doublets. 
When the spin-orbit coupling is weaker than the CEF splitting, the main components are approximately $c_{1yz}\sim 1/\sqrt{2}$, $c_{2yz}\sim 1/\sqrt{2}$, $c_{3v}\sim 1$, $c_{4u}\sim 1$, and $c_{5xy}\sim 1$. 
The subscripts $a$ and $b$ denote the two spin states within each Kramers doublet. 
Among these states, $d_1$ and $d_4$ belong to the $\Gamma_6$ representation, while $d_2$, $d_3$, and $d_5$ belong to the $\Gamma_7$ representation.  

We set the tetragonal point-charge configuration for the Ce-$5d$ CEF following Yamada \textit{et al.} [11], such that $\varepsilon_{yz}=\varepsilon_{zx}=-348$, $\varepsilon_v=-264$, $\varepsilon_u=-107$, and $\varepsilon_{xy}=1067$ meV. The spin-orbit coupling constant was taken as $\zeta_{5d}=70$ meV [11], which yields $\varepsilon_{d1}=-407$, $\varepsilon_{d2}=-348$, $\varepsilon_{d3}=-234$, $\varepsilon_{d4}=-82.2$, and $\varepsilon_{d5}=1072$ meV.  The coefficients in (\ref{eqS:dcefSO}) are $(c_{1yz}, c_{1u})=(0.68, 0.27)$, $(c_{2yz}, c_{2v}, c_{2xy})=(0.59, 0.54, 0.056)$, $(c_{3yz}, c_{3v}, c_{3xy})=(0.38, 0.84, 0.025)$, $(c_{4yz}, c_{4u})=(0.19, 0.96)$, and $(c_{5yz}, c_{5v}, c_{5xy})=(0.024, 0.051, 0.998)$. 

\newpage
\begin{center}
\textbf{B. Ce-$4f$ states in the crystal field}
\end{center}

From Nikitin \textit{et al.} [12], the CEF eigenstates of Ce-$4f$ are written in the $|J=5/2, M\rangle$ ($M=5/2, 3/2, \cdots, -5/2$) basis as
\begin{align}
|f_{7(1)a}\rangle &= 0.306 |5/2\rangle - 0.952|-3/2\rangle \;,  \nonumber \\
|f_{7(1)b}\rangle &= -0.952 |3/2\rangle + 0.306|-5/2\rangle \;, \nonumber \\
|f_{7(2)a}\rangle &= 0.306 |3/2\rangle + 0.952|-5/2\rangle \;,  \label{eqS:fcefJ}\\
|f_{7(2)b}\rangle &= 0.952 |5/2\rangle + 0.306|-3/2\rangle \;, \nonumber \\
|f_{6a}\rangle &=  |1/2\rangle \;, \nonumber \\
|f_{6b}\rangle &=  |-1/2\rangle \;. \nonumber 
\end{align}
The subscripts $7(1)$, $7(2)$, and 6 represent the $\Gamma_{7}^{(1)}$, $\Gamma_{7}^{(2)}$, and $\Gamma_{6}$ states, respectively. 
The corresponding energy levels are $\varepsilon_{7(1)}^{(0)} = 0$, 
$\varepsilon_{7(2)}^{(0)} = 10.78$, and $\varepsilon_{6}^{(0)} = 14.26 $ meV. 

These states can be rewritten using the cubic harmonic functions for the $f$ electrons with $l=3$. 
\begin{align}
|f_{7(1)a}\rangle &= -0.651i |f_{xyz,\uparrow}\rangle + 0.487 |f_{\beta z,\uparrow}\rangle +  0.0622 (|f_{\alpha x,\downarrow} \rangle -i |f_{\alpha y,\downarrow} \rangle) + 0.407 (|f_{\beta x,\downarrow} \rangle +i |f_{\beta y,\downarrow} \rangle ) \;, \nonumber \\
|f_{7(1)b}\rangle &= 0.0622 (|f_{\alpha x,\uparrow} \rangle +i |f_{\alpha y,\uparrow} \rangle) + 0.407 (|f_{\beta x,\uparrow} \rangle -i |f_{\beta y,\uparrow} \rangle )  -0.651i |f_{xyz,\downarrow}\rangle - 0.487 |f_{\beta z,\downarrow}\rangle  \;, \nonumber \\
|f_{7(2)a}\rangle &= -0.564 (|f_{\alpha x,\uparrow} \rangle +i |f_{\alpha y,\uparrow} \rangle) + 0.290 (|f_{\beta x,\uparrow} \rangle -i |f_{\beta y,\uparrow} \rangle )  -0.0718 |f_{xyz,\downarrow}\rangle +0.437 |f_{\beta z,\downarrow}\rangle  \label{eqS:dcefCub}\\
|f_{7(2)b}\rangle &= -0.0718i |f_{xyz,\uparrow}\rangle - 0.437 |f_{\beta z,\uparrow}\rangle - 0.564 (|f_{\alpha x,\downarrow} \rangle -i |f_{\alpha y,\downarrow} \rangle) + 0.290 (|f_{\beta x,\downarrow} \rangle +i |f_{\beta y,\downarrow} \rangle )  \;, \nonumber \\
|f_{6a}\rangle &= 0.327 (|f_{\alpha x,\uparrow} \rangle -i |f_{\alpha y,\uparrow} \rangle) + 0.423 (|f_{\beta x,\uparrow} \rangle +i |f_{\beta y,\uparrow} \rangle )  + 0.655 |f_{\alpha z,\downarrow}\rangle  \;, \nonumber \\
|f_{6a}\rangle &= -0.655 |f_{\alpha z,\uparrow}\rangle + 0.327 (|f_{\alpha x,\downarrow} \rangle +i |f_{\alpha y,\downarrow} \rangle) + 0.423 (|f_{\beta x,\downarrow} \rangle -i |f_{\beta y,\downarrow} \rangle )  \;. \nonumber 
\end{align}

The cubic harmonics for $f$-electrons are expressed in the $|l=3, m\rangle$ ($m=3,2,\cdots,-3$) basis as
\begin{align}
|f_{xyz}\rangle &= -\frac{i}{\sqrt{2}}(|3,2\rangle - |3,\bar{2}\rangle) \;, \nonumber \\
|f_{\alpha x} \rangle &= \frac{1}{4}(-\sqrt{5}|3,3\rangle + \sqrt{3}|3,1\rangle - \sqrt{3}|3,\bar{1}\rangle + \sqrt{5}|3,\bar{3}\rangle) \;,  \nonumber \\
|f_{\alpha y} \rangle &= -\frac{i}{4}(\sqrt{5}|3,3\rangle + \sqrt{3}|3,1\rangle + \sqrt{3}|3,\bar{1}\rangle + \sqrt{5}|3,\bar{3}\rangle) \;,  \nonumber \\
|f_{\alpha z} \rangle &= |3,0\rangle \;,   \label{eqS:fCub}\\
|f_{\beta x} \rangle &= \frac{1}{4}(\sqrt{3}|3,3\rangle + \sqrt{5}|3,1\rangle - \sqrt{5}|3,\bar{1}\rangle - \sqrt{3}|3,\bar{3}\rangle) \;,  \nonumber \\
|f_{\beta y} \rangle &= -\frac{i}{4}(\sqrt{3}|3,3\rangle - \sqrt{5}|3,1\rangle - \sqrt{5}|3,\bar{1}\rangle + \sqrt{3}|3,\bar{3}\rangle) \;,  \nonumber \\
|f_{\beta z} \rangle &= \frac{1}{\sqrt{2}}(|3,2\rangle + |3,\bar{2}\rangle) \;.  \nonumber
\end{align}

\newpage
\begin{center}
\textbf{C. On-site $4f$-$5d$ hybridization due to noncentrosymmetry}
\end{center}

In CeCoSi, a local electric-dipole field $E$ along the $z$ direction exists in the noncentrosymmetric CEF. 
When the matrix element between the $4f$-state $|3,m'\rangle$ and the $5d$-state $|2,m\rangle$ is evaluated through the electric field potential, the following matrix element does not vanish:  
\begin{align}
\langle 4f,3,m' | eE z| 5d, 2,m \rangle &= eE \int_0^{\infty} r^3 R_{4f}^{\;*}(r)R_{5d}(r) dr \int_0^{\pi} d\theta \int_0^{2\pi} Y_{3m}^{\;*}(\theta,\phi)Y_{2m}(\theta,\phi)  \cos\theta \sin\theta d\phi \;.
\label{eqS:fdinteg}
\end{align}
This non-vanishing matrix element allows for on-site $4f$-$5d$ hybridization. 
The angular part of the matrix elements for the $E_z$ field is calculated as 
\begin{align}
\mathcal{H}_{fd,z} &= \bordermatrix{
                         &  |f_{xyz} \rangle   &  |f_{\alpha x} \rangle  &   |f_{\alpha y} \rangle  &  |f_{\alpha z} \rangle  &   |f_{\beta x} \rangle  &   |f_{\beta y} \rangle  &   |f_{\beta z} \rangle \cr
    \langle d_u |  &  0 & 0 & 0 & \frac{3}{\sqrt{35}} & 0 & 0 & 0 \cr
    \langle d_v |  & 0 & 0 & 0 & 0 & 0 & 0 & \frac{1}{\sqrt{7}} \cr
    \langle d_{yz} |  & 0 & 0 & -\sqrt{\frac{3}{35}} & 0 & 0 & \frac{1}{\sqrt{7}} & 0 \cr
    \langle d_{zx} |  & 0 & -\sqrt{\frac{3}{35}} & 0 & 0 & -\frac{1}{\sqrt{7}} & 0 & 0 \cr
    \langle d_{xy} |  & \frac{1}{\sqrt{7}} & 0 & 0 & 0 & 0 & 0 & 0 \cr
    } \;.  \label{eqS:HdfcubEz}
\end{align}
Then, the $f$-$d$ hybrid Hamiltonian, through the local electric-dipole field $E_z$ in the tetragonal CEF, is given by 
\begin{align}
\mathcal{H}_{fd} &= \bordermatrix{
    &  |d_{1a} \rangle   &  |d_{1b} \rangle  &   |d_{2a} \rangle  &  |d_{2b} \rangle  &  |d_{3a} \rangle  &  |d_{3b} \rangle  &   |d_{4a} \rangle  &  |d_{4b} \rangle  &   |d_{5a} \rangle  &  |d_{5b} \rangle    \cr
\langle f_{7(1)a} |  &  0 & 0 & A & 0 & -B & 0 & 0 & 0 & 0 & C \cr
\langle f_{7(1)b} |  &  0 & 0 & 0 & -A & 0 & B & 0 & 0 & -C & 0 \cr
\langle f_{7(2)a} |  &  0 & 0 & 0 & D & 0 & -E & 0 & 0 & -F & 0 \cr
\langle f_{7(2)b} |  &  0 & 0 & -D & 0 & E & 0 & 0 & 0 & 0 & F \cr
\langle f_{6a} |  &  G & 0 & 0 & 0 & 0 & 0 & -H & 0 & 0 & 0 \cr
\langle f_{6b} |  &  0 & G & 0 & 0 & 0 & 0 & 0 & -H & 0 & 0 \cr
} \;. \label{eq:S10}
\end{align}
The constant parameters $A$, $B$, $C$, $D$, $E$, $F$, $G$, and $H$ depend on the spin-orbit-coupling constant and the CEF level scheme of the $5d$ orbital. 
It is important to note that there is no matrix element between $4f$-$\Gamma_7$ and $5d$-$\Gamma_6$ ($d_1$ and $d_4$), and no matrix element between $4f$-$\Gamma_6$ and $5d$-$\Gamma_7$ ($d_2$, $d_3$, and $d_5$). 
Through the mixing terms $\langle f_{7(1)} | \mathcal{H}_{fd} | d_2 \rangle$, $\langle f_{7(1)} | \mathcal{H}_{fd} | d_3 \rangle$, and $\langle f_{7(1)} | \mathcal{H}_{fd} | d_5 \rangle$, the energy level of $4f$-$\Gamma_7^{(1)}$ is lowered through second-order perturbation, assuming the $5d$ level lies at a higher energy than the $4f$ level. The same lowering of energy occurs for $4f$-$\Gamma_7^{(2)}$ and $4f$-$\Gamma_6$. 
This should be the actual CEF level scheme experimentally confirmed by inelastic neutron scattering [12]. 
By taking the $f$-$d$ hybridization constant in (\ref{eqS:fdinteg}) as 100 meV [11], we obtain $A=32$, $B=2.9$, $C=23$, $D=15$, $E=9.6$, $F=3.8$, $G=44$, and $H=22$ meV. 
The bare $d$ level was set at 500 meV above the $f$ level.

The essence of the energy lowering due to the monoclinic distortion lies in the appearance of the matrix element 
$\langle f_{7(1)} | \mathcal{H}_{fd} | d_1 \rangle$, which arises because the distortion admixes the $d_2$, $d_3$, and $d_5$ states of $\Gamma_7$ symmetry into the $d_1$ state of $\Gamma_6$ symmetry. Furthermore, when the electric field tilt away from the $c$-axis, $E_x$ and $E_y$ components appear at the Ce site, which further enhance this matrix element. These factors contribute to an additional lowering of the $4f$ energy level, as illustrated in Fig.~\ref{fig:dfmixing}(a). 

To estimate the energy gain due to the distortion, we shifted the positions of the point charges and changed the Ce-$5d$ CEF symmetry to monoclinic, which split the degenerate $d_{yz}$ and $d_{zx}$ states by 20 meV. 
This distortion effectively corresponds to a tilt of the $c$-axis by $1.5^{\circ}$. 
After including the spin-orbit coupling of $\zeta_{5d}=70$ meV, the energies are modified to $\varepsilon_{d1}=-422$, $\varepsilon_{d2}=-355$, $\varepsilon_{d3}=-226$, $\varepsilon_{d4}=-72.1$ meV. The $d_{xy}$ state at high energy was neglected.  
Finally, tilting the local electric field by $10^{\circ}$ from the $c$-axis, which may effectively be realized by a horizontal displacement of the Ce atom, lowers the energy of the lowest $\Gamma_{7(1)}$ state by 1.5 meV relative to the tetragonal state. 
This estimate explains how the transition temperature $T_0 = 12$ K can arise from the $f$-$d$ hybridization mechanism.  

\newpage
\begin{center}
\textbf{D. The simplest model including $f$-$d$ hybridization and $d$-$d$ transfer}
\end{center}

\begin{figure}[b]
\begin{center}
\includegraphics[width=5cm]{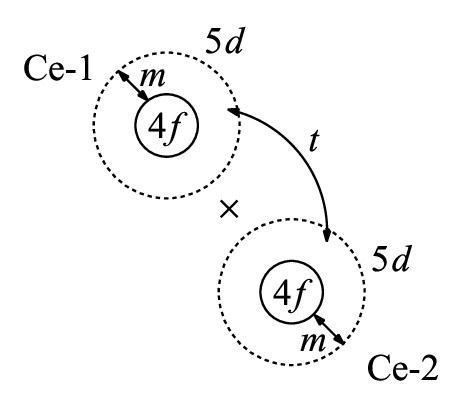}
\end{center}
\caption{
Schematic illustration of the simplest model including $f$-$d$ hybridization and $d$-$d$ transfer. The cross mark represents an inversion center. 
}
\label{fig:fdmixddtrans}
\end{figure}

Let us consider a simple model with one orbital state each for $4f$ and $5d$, for example, $\Gamma_7^{(1)}$ and $d_1$, respectively, and consider two Ce atoms, Ce-1 and Ce-2. A schematic illustration is shown in Fig.~\ref{fig:fdmixddtrans}.
If we assume the energy levels of the $4f$ and $5d$ states at 0 and $\Delta$, respectively, with $f$-$d$ mixing matrix $m$, and $d$-$d$ transfer $t$, the two-atom Hamiltonian for spin state $a$ is expressed as 
\begin{align}
\mathcal{H} &= \bordermatrix{
                         &  |f_{a}(1) \rangle   &  |f_{a}(2) \rangle  &   |d_{a}(1) \rangle  &  |d_{a}(2) \rangle  \cr
    \langle f_{a}(1) |  &  0 & 0 & m &  0 \cr
    \langle f_{a}(2)|  & 0 & 0 & 0 & -m \cr
    \langle d_{a}(1) |  & m & 0 & \Delta & t  \cr
    \langle d_{a}(2) |  & 0 & -m& t & \Delta \cr
    }\;.  \label{eq:S11}
\end{align}
The sign of the $f$-$d$ mixing is opposite for Ce-1 and Ce-2 because the directions of the electric field is opposite. 
The same Hamiltonian applies for the spin state $b$. 

The resultant levels are four Kramers doublets, with energies given by
\begin{align}
\varepsilon_{fa} &= \varepsilon_{fb} = -\frac{m^2}{\Delta - t} \; \nonumber \\
\varepsilon_{fa}^{\;*}  &= \varepsilon_{fb}^{\;*}  = -\frac{m^2}{\Delta + t} \; \nonumber\\
\varepsilon_{da}  &= \varepsilon_{db} =  \Delta - t + \frac{m^2}{\Delta - t} \; \label{eq:S12} \\
\varepsilon_{da}^{\;*}  &= \varepsilon_{db}^{\;*}  = \Delta + t + \frac{m^2}{\Delta + t} \;. \nonumber
\end{align}
If $\Delta \gg t$, the $4f$ energy level is lowered due to the hybridization. 
This is the schematic level scheme shown in Fig.~\ref{fig:dfmixing}(a) in the main text. 
The splitting between $\varepsilon_{da}$ and $\varepsilon_{da}^{\;*}$, which is $\sim 2t$, reflects the bonding and anti-bonding nature of the $5d$ orbitals between Ce-1 and Ce-2. 

The eigenfunctions of these $f$-$d$ hybrid states are
\begin{align}
\varphi_{fa} &= \alpha\{ |f_{a}(1) \rangle - |f_{a}(2) \rangle \} + \beta \{ |d_{a}(1) \rangle + |d_{a}(2) \rangle  \} \;, \nonumber  \\
\varphi_{fa}^{\;*} &= \alpha\{|f_{a}(1) \rangle + |f_{a}(2) \rangle \} + \beta \{ |d_{a}(1) \rangle - |d_{a}(2) \rangle \}  \;.  \label{eq:S13}
\end{align}
Here, $\alpha$ and $\beta$ are constants, with $\alpha \gg \beta$. The expressions for $\varphi_{fb}$ and $\varphi_{fb}^{\;*}$ are the same. 
In $\varphi_{fa}$, $d_a(1)$ and $d_a(2)$ have the same signs, forming a bonding orbital. 
It should be noted that the splitting between $\varepsilon_{fa}$ and $\varepsilon_{fa}^{\;*}$ ($\varepsilon_{fa} < \varepsilon_{fa}^{\;*}$) is in practice very small. This is because it originates from neglecting the electronic correlation $U$ between two electrons occupying the same $f$ orbital, which essentially suppresses the $f^2$ state.

\newpage
\begin{center}
\textbf{E. Curie-type softening of the elastic modulus due to Ce-Ce dimer formation }
\end{center}

When an atom is not located at a centrosymmetric site, a bilinear coupling term 
\begin{equation}
E_{\text{odd}} =\sum_{i,\alpha}\lambda_{i\alpha}e_{i}u_{\alpha}\;\;\;\; (i=1\sim 6, \alpha=x,y,z)
\end{equation} 
is allowed. 
The uniform deformations are defined as $e_1=e_{xx}$, $e_2=e_{yy}$, $e_3=e_{zz}$, $e_4=e_{yz}$, $e_5=e_{zx}$, and $e_6=e_{xy}$. 
The displacement $\bm{u}$ represents the shift of the equilibrium position of the atom. 
The general expression of this odd-parity term for the $4mm$ point symmetry ($2c$ site of Ce in the space group $P4/nmm$) is 
\begin{align}
E_{\text{odd}} &= -\lambda_{1z} (e_1+e_2) u_z - \lambda_{3z} e_3 u_z  - \lambda_{4y} (e_4 u_y + e_5 u_x) \;.
\label{eq:Eodd}
\end{align}
In fourfold symmetry, the $e_6$ term does not appear in the first-order expression. 
A schematic illustration of this effect for Ce-1 and Ce-2 at $(0.25, 0.25, z)$ and $(0.75, 0.75, 1-z)$, respectively, is shown in Fig.~\ref{fig:displacement}.

\begin{figure}[b]
\begin{center}
\includegraphics[width=9cm]{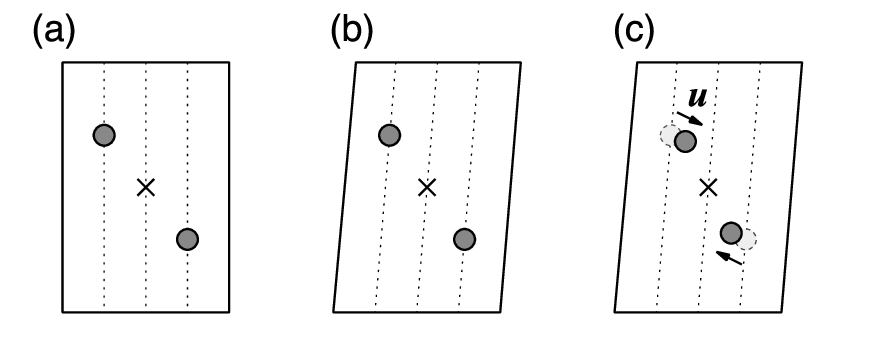}
\end{center}
\caption{
Schematic illustration of the mutual relationship between strain and atomic displacements.  
(a) Original positions of Ce-1 and Ce-2 at $(0.25, 0.25, z)$ and $(0.75, 0.75, 1-z)$, respectively,  in the tetragonal lattice. 
(b) Uniform atomic displacements corresponding to the uniform strain. 
(c) Shift of the equilibrium positions induced by uniform strain, and conversely, strain induced by atomic displacements via the bilinear coupling. 
}
\label{fig:displacement}
\end{figure}

When a monoclinic distortion occurs such that the $c$-axis tilts to the [110] direction, one of the four nearest-neighbor Ce-1 Ce-2 pairs forms the state expressed by Eq.~(\ref{eq:S13}). This pairing phenomenon may be regarded as dimerization, driven by the energy gain associated with the monoclinic distortion as explained above. There are four possible parings. 
Assuming that Ce-1 is located at $(0.25, 0.25, z)$, the four corresponding Ce-2 atoms are at A: $(0.75, 0.75, 1-z)$, B: $(0.75, -0.25, 1-z)$, C: $(-0.25, -0.25, 1-z)$, and D: $(-0.25, 0.75, 1-z)$. 
In the tetragonal phase, the probability of forming a dimer is equal for all four pairs. 

Let us assume that when Ce-1 forms a dimer, the equilibrium position shifts to $(0.25+u_a, 0.25+u_a, z+u_c)$ for pair A, 
$(0.25+u_a, 0.25-u_a, z+u_c)$ for pair B, $(0.25-u_a, 0.25-u_a, z+u_c)$ for pair C, and $(0.25-u_a, 0.25+u_a, z+u_c)$ for pair D. 
The corresponding Ce-2 atom shifts in the opposite direction so that inversion symmetry is preserved at the center of the pair (Fig.~\ref{fig:displacement}). 
These displacements are actually predicted by the first-principles ``free-f" calculation (Fig.~\ref{fig:DispItinerant}).
The energies of Ce-1 for the four possible dimer configurations are given by:  
\begin{align}
E_{\text{A}} &= E_0 -\lambda_{1z}(e_1 + e_2) u_c - \lambda_{3z} e_3 u_c - \lambda_{4y} (e_4 + e_5) u_a  \;, \nonumber \\
E_{\text{B}} &= E_0 -\lambda_{1z}(e_1 + e_2) u_c - \lambda_{3z} e_3 u_c - \lambda_{4y} (-e_4 + e_5) u_a \;,  \nonumber \\
E_{\text{C}} &= E_0 -\lambda_{1z}(e_1 + e_2) u_c - \lambda_{3z} e_3 u_c - \lambda_{4y} (-e_4 - e_5) u_a  \;, \label{eq:S14}\\
E_{\text{D}} &= E_0 -\lambda_{1z}(e_1 + e_2) u_c - \lambda_{3z} e_3 u_c - \lambda_{4y} (e_4 - e_5) u_a  \;.  \nonumber
\end{align}
The same expressions apply to Ce-2. 

The probability of forming a dimer in the $\gamma$ pair ($\gamma$=A, B, C, D) at temperature $T$ is given by
\begin{equation}
p_{\gamma} = \frac{e^{-E_{\gamma}/\kB T}}{\sum_{\nu} e^{-E_{\nu}/\kB T} } \;. \label{eq:S15}
\end{equation}
If we assume $\kB T \gg \lambda e u$, the following probabilities are obtained.  
\begin{align}
p_{\text{A}} &= \frac{1}{4} \left\{ 1+ \frac{ \lambda_{4y} (e_4 + e_5) u_a}{\kB T} \right\}  \;, \nonumber \\
p_{\text{B}} &= \frac{1}{4} \left\{ 1+ \frac{ \lambda_{4y} (-e_4 + e_5) u_a}{\kB T} \right\} \;, \nonumber \\
p_{\text{C}} &= \frac{1}{4} \left\{ 1+ \frac{ \lambda_{4y} (-e_4 - e_5) u_a}{\kB T} \right\} \;, \\
p_{\text{D}} &= \frac{1}{4} \left\{ 1+ \frac{ \lambda_{4y} (e_4 - e_5) u_a}{\kB T} \right\} \;. \nonumber 
\end{align}
The thermal average of $E_{\text{odd}}$, representing the dimerization energy for each Ce atom in this context, is   
\begin{align}
E_{\text{odd}} &= \sum_{\gamma} p_{\gamma} E_{\gamma} \nonumber \\
 &= E_0 - \lambda_{1z} (e_1 + e_2 ) u_c - \lambda_{3z} e_3 u_c -\frac{\lambda_{4y}^{\;2} (e_4^{\;2} + e_5^{\;2}) u_a^{\;2}}{\kB T} \;.
\end{align}
Here, the last term leads to a Curie-type softening, whereas the $(e_1 + e_2)u_c$ and $e_3 u_c$ terms produce a uniform lattice distortion, as explained below.

The elastic energy under uniform strain is expressed as
$ E_{\text{el}} = \frac{1}{2} \sum_{ij} C_{ij} e_i e_j$, 
where $C_{ij}$ denote the elastic stiffness constants and $e_i$ the strain components. 
In a tetragonal symmetry this reduces to 
\begin{align}
E_{\text{el}} &= \frac{1}{2}C_{11} (e_1^{\;2} + e_2^{\;2} ) + \frac{1}{2}C_{33} e_3^{\;2} + \frac{1}{2} C_{44} (e_4^{\;2} + e_5^{\;2} ) + \frac{1}{2}C_{66} e_6^{\;2} 
\nonumber \\
 &\;\;  + C_{13} (e_2 e_3 + e_3 e_1 ) + C_{12} e_1 e_2 \;.
 \label{eq:Eeltet}
\end{align}
Therefore, the total energy, given by $E = E_{\text{el}} +  NE_{\text{odd}}$ (where $N$ denotes the number of Ce atoms per unit volume), can be written as 
\begin{align}
E &= \frac{1}{2}C_{11} (e_1^{\;2} + e_2^{\;2} ) + \frac{1}{2}C_{33} e_3^{\;2} + \frac{1}{2} \Bigl(C_{44} -\frac{ 2N \lambda_{4y}^{\;2} u_a^{\;2}}{\kB T}  \Bigr) (e_4^{\;2} + e_5^{\;2} ) + \frac{1}{2}C_{66} e_6^{\;2} 
\nonumber \\
 &\;\;  + C_{13} (e_2 e_3 + e_3 e_1 ) + C_{12} e_1 e_2 
 + N \{ E_0 - \lambda_{1z} (e_1 + e_2 ) u_c - \lambda_{3z} e_3 u_c \} \;.
\end{align}
The $(e_4^{\;2} + e_5^{\;2})$ term directly indicates that the elastic modulus $C_{44}$ exhibits a Curie-type softening. 
The $(e_1 + e_2)u_c$ and $e_3 u_c$ terms give rise to a uniform distortion. 
From the equilibrium conditions, 
\begin{equation}
\frac{\partial E}{\partial e_{1}} = \frac{\partial E}{\partial e_{2}}  = \frac{\partial E}{\partial e_{3}}   = 0 \;, 
\end{equation}
we obtain 
\begin{align}
 e_1 &= e_2 =  \frac{ N( C_{13} \lambda_{3z} - C_{33} \lambda_{1z} ) u_c}{2C_{13}^{\;2} - C_{33}(C_{11} + C_{12})}  \;, \nonumber \\
 e_3 &=  \frac{ N\{ 2C_{13} \lambda_{1z} - (C_{11} + C_{12}) \lambda_{3z} \} u_c}{2C_{13}^{\;2} - C_{33}(C_{11} + C_{12})}  \;. 
\end{align}
Thus, the dimer formation in the present model induces a change in the lattice parameters $a$ and $c$.

Importantly, the quadrupolar degeneracy of the single-ion CEF ground state is not required here. 
$C_{44}$ corresponds to the $\Delta_{3,z}$, $\Sigma_{4,z}$, and $\Lambda_{5,xy}$ modes, all of which exhibit softening in the present measurement.   
The temperature-dependent softening in other modes, such as $C_{11}$, $C_{33}$, $C_{12}$, $C_{13}$, and $C_{66}$, may arise from the even-parity term, which corresponds to the conventional quadrupole-strain coupling. This effect, however, is van-Vleck type and is expected to be weak in CeCoSi owing to the well-separated Kramers doublets. 

\vspace{5mm}
\noindent
\newpage
\textbf{references} \\
\renewcommand{\labelenumi}{[\arabic{enumi}]}
\begin{enumerate}
\item X. Gonze, B. Amadon, G. Antonius, F. Arnardi, L. Baguet, J.-M. Beuken, J. Bieder, F. Bottin, J. Bouchet, E. Bousquet, N. Brouwer, F. Bruneval, G. Brunin, T. Cavignac, J.-B. Charraud, W. Chen, M. C\^{o}t\'{e}, S. Cottenier, J. Denier, G. Geneste, P. Ghosez, M. Giantomassi, Y. Gillet, O. Gingras, D. R. Hamann, G. Hautier, X. He, N. Helbig, N. Holzwarth, Y. Jia, F. Jollet, W. Lafargue-Dit-Hauret, K. Lejaeghere, M. A. L. Marques, A. Martin, C. Martins, H. P. C. Miranda, F. Naccarato, K. Persson, G. Petretto, V. Planes, Y. Pouillon, S. Prokhorenko, F. Ricci, G.-M. Rignanese, A. H. Romero, M. M. Schmitt, M. Torrent, M. J. van Setten, B. V. Troeye, M. J. Verstraete, G. Z\'{e}rah, J. W. Zwanziger, Comput. Phys. Commun. \textbf{248}, 107042 (2020).
\item A. H. Romero, D. C. Allan, B. Amadon, G. Antonius, T. Applencourt, L. Baguet, J. Bieder, F. Bottin, J. Bouchet, E. Bousquet, F. Bruneval, G. Brunin, D. Caliste, M. C\^{o}t\'{e}, J. Denier, C. Dreyer, P. Ghosez, M. Giantomassi, Y. Gillet, O. Gingras, D. R. Hamann, G. Hautier, F.Jollet, G. Jomard, A. Martin, H. P. C. Miranda, F. Naccarato, G. Petretto, N. A. Pike, V. Planes, S. Prokhorenko, T. Rangel, F. Ricci, G.-M. Rignanese, M. Royo, M. Stengel, M. Torrent, M. J. van Setten, B. V. Troeye, M. J. Verstraete, J. Wiktor, J. W. Zwanziger, X. Gonze, J. Chem. Phys. \textbf{152}, 124102 (2020).
\item P. E. Bl\"{o}chl, Phys. Rev. B \textbf{50}, 17953 (1994). 
\item M. Torrent, F. Jollet, F. Bottin, G. Z\'{e}rah, X. Gonze, Comput. Mater. Sci. \textbf{42}, 337 (2008).
\item J. P. Perdew and Y. Wang, Phys. Rev. B \textbf{45}, 13244 (1992).
\item H. J. Monkhorst and J. D. Pack, Phys. Rev. B \textbf{13}, 5188 (1976).
\item B. G. Pfrommer, M. C\^{o}t\'{e}, S. G. Louie, and M. L. Cohen, J. Comput. Phys. \textbf{131}, 233 (1997).
\item X. Gonze, Phys. Rev. B \textbf{55}, 10337 (1997).
\item X. Gonze and C. Lee, Phys. Rev. B \textbf{55}, 10355 (1997).
\item N. A. W. Holzwarth, A. R. Tackett, G. E. Matthews, Comput. Phys. Commun., \textbf{135}, 329 (2001).
\item T. Yamada, Y. Yanagi, and K. Mitsumoto, J. Phys. Soc. Jpn. \textbf{93}, 053703 (2024). 
\item S. E. Nikitin, D. G. Franco, J. Kwon, R. Bewley, A. Podlesnyak, A. Hoser, M. M. Koza, C. Geibel, and O. Stockert, Phys. Rev. B \textbf{101}, 214426 (2020). 
\end{enumerate}

\end{widetext}

\end{document}